\documentclass{mcom-l}
\usepackage{fancyhdr}
\usepackage{enumitem}
\usepackage{float}
\usepackage{placeins}
\usepackage{graphicx}
\usepackage{hyperref}

\copyrightinfo{}{}

\newtheorem{theorem}{Theorem}[section]

\theoremstyle{definition}

\theoremstyle{remark}

\numberwithin{equation}{section}

\begin{document}

\newcommand{\SubItem}[1]{
    {\setlength\itemindent{15pt} \item[-] #1}
}

\title{Analysis of a Spotify Collaboration Network for Small-world Properties}

\author{Raquel Ana Magalhães Bush}
\address{Department of Mathematics, University of Massachusetts Dartmouth, North Dartmouth, Massachusetts 02747}
\curraddr{}
\email{rbush@umassd.edu}
\thanks{}

\subjclass[2020]{Primary 05C82, 91D30; Secondary 05C90, 05C75}

\date{}

\dedicatory{}

\keywords{Small-world networks, artist collaboration networks}

\begin{abstract}
This paper examines the small-world properties of a Spotify artist feature collaboration network, focusing on clustering coefficients and network diameter. I analyze the giant component and subgraphs based on genres, country-specific charts, and detected communities to assess their small-world characteristics. Results indicate that the network is scale-free and follows a power-law degree distribution, with highly popular artists serving as central hubs. Louvain community detection reveals distinct collaboration clusters aligned with genre-based and industry-driven connections. These findings offer insights into music recommendation systems and digital collaboration trends, contributing to a broader understanding of artist networks in the digital age.
\end{abstract}

\maketitle

\section*{Introduction}
In recent years, the study of networks and their graphs has extended into diverse fields, providing insights into the structures and behaviors of interconnected systems and social networks. One domain where network analysis is proving increasingly valuable is the study of music collaboration networks. Much like social networks, these reveal intricate patterns of connection and influence. Exploring these networks can provide unique insights into how creativity and collaboration thrive in the digital era. 

This paper analyzes a network involving artists who have charted on Spotify and their collaborators in features, focusing on identifying and exploring small-world properties within the network. Small-world networks are defined by their higher clustering and smaller diameters compared to random graphs and regular lattices of the same size and density. These traits are commonly found in social networks and collaboration systems, and they provide an ideal framework for understanding artist collaboration dynamics.

Using a dataset of artists and collaborations among those who appeared in Spotify's weekly charts, this study constructs a graph of approximately 156,000 artists with over 300,000 collaborative connections \cite{F}. Artists are represented as vertices, and feature collaborations are treated as edges between these vertices. By analyzing the network’s giant component, this research seeks to determine whether this collaboration network exhibits small-world properties. All analysis was performed using Python, with the NetworkX library and other standard packages for network analysis. The Cytoscape application was used to create visualizations of the network and one of the subgraphs. All other visuals were created solely using Python.

As part of this analysis, clustering and diameter are calculated and compared to those of the Erdős–Rényi $G(n,p)$ random graphs, as well as to the Wattz-Strogratz graphs with $p = 0$ \cite{E}\cite{Wa}. Comparable random graphs and regular lattices have both equal size and density. Additional information, such as edge distribution, are gathered and examined. Genre-specific subgraphs are created to examine how collaboration patterns differ across musical genres, which offers insights into genre-specific tendencies in artist collaborations. The Louvain method, which efficiently detects densely clustered communities in large, complex networks, is applied. Furthermore, this study examines the collaboration patterns of chart-topping artists both globally and within specific countries, including the United States, United Kingdom, India, Brazil, South Africa, and Japan, to explore how small-world properties manifest across diverse regional and global music networks.

This analysis contributes to the growing body of research on music networks by providing a structural perspective on artist collaborations within a digital music platform. The findings provide a deeper understanding of the dynamics that drive musical creativity and innovation. For instance, hubs such as globally renowned artists like Steve Aoki play a pivotal role in maintaining connectivity, while tightly knit genre-specific communities highlight unique collaboration trends. 

This research not only contributes to the growing field of music network analysis but also offers practical applications for platforms like Spotify. Findings from this study may offer new ways to understand the impact of genre on collaboration frequency and clustering, as well as provide a basis for future research into network-driven music recommendation systems or community-based
artist discovery algorithms. In doing so, it bridges the gap between theoretical network analysis and its implications for the modern music industry.

\section*{Related Work}
The existing body of work shows that the analysis of small-world properties in music networks may have significant implications for understanding collaboration, creativity, and community structure.

Uzzi and Spiro’s study on Broadway musical creators demonstrates that small-world networks foster an optimal balance of connectivity and cohesion. They found a parabolic relationship between high clustering and small diameter and creative success, where moderate small-world qualities promoted creativity, but excessive connectivity led to homogenization and diminished innovation \cite{U}.

Expanding on this, Park et al. explored the network of classical composers and confirmed that it exhibits small-world properties. They found that this network structure facilitates efficient information flow and maintains diversity across periods and styles. Notably, the authors observed that prominent composers are more likely to attract additional collaborations, which intensifies the small-world effect \cite{P}.

Similarly, Teitelbaum et al. investigated community structures within collaboration and similarity networks in the music industry, and they found that these networks inherently possess small-world characteristics, which drive the formation of genre-based communities. They highlight how genre, geography, and generational factors play roles in shaping dense clusters with frequent interconnections, revealing the strong social and musical dynamics underpinning these communities \cite{T}.

These prior studies on music collaboration networks share a focus on connections formed through in-person or tangible interactions, and they primarily represent the collaborative dynamics within geographic areas and the era before the rise of online music platforms. These studies provide a comprehensive understanding of musical collaboration networks in traditional contexts, but they do not fully address the evolving nature of collaboration in the digital age. In the last few years, online platforms such as Spotify and the Internet at large have become a dominant force in shaping musical collaborations. Digital collaborations do not require simultaneous physical presence, allowing artists to work asynchronously and at a lower cost. This could increase the frequency and geographical diversity of artist collaborations, leading to more interconnected vertices and higher clustering within specific sub-communities.

In 2021, South et al. studied eigenvector centrality and its critical transitions in a Spotify collaboration network of over one million artists, and they highlighted a major structural shift from classical to rap artists at certain popularity thresholds \cite{S}. 

While their research focuses on centrality metrics and how popularity bias impacts the network, this paper takes a different approach by examining small-world properties like clustering and diameter. By investigating genre- and region-specific subgraphs, it offers new insights into collaboration patterns, emphasizing how genre-based modularity and key hubs contribute to the network's overall connectivity.

\section*{The Entire Graph}
\subsection*{Overview of the Spotify Collaboration Network}
\begin{figure}[H]
    \centering
    \includegraphics[width=1\linewidth]{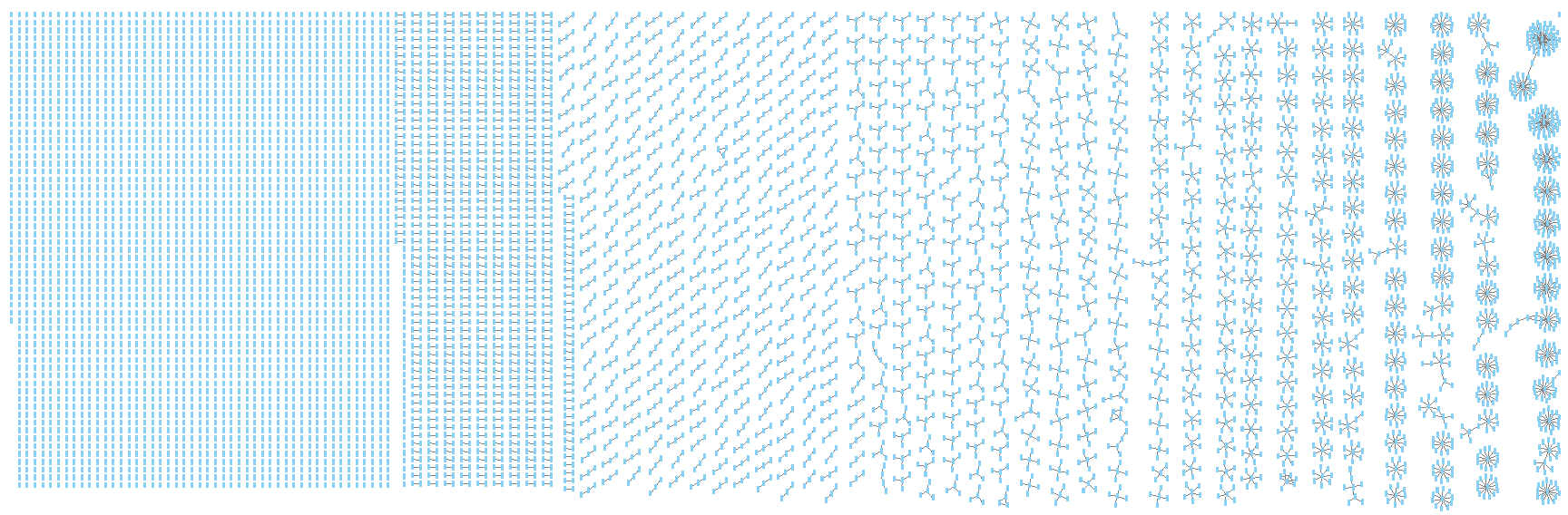}
    \caption{A visualization of the disconnected graph.}
    \label{clusters}
\end{figure}

The Spotify collaboration network, shown in Figure~\ref{clusters}, consists of 156,326 vertices (artists) connected by 300,386 edges (collaborations). The primary focus of this analysis is the giant component of the graph, which encompasses 148,386 vertices and accounts for the vast majority of the network. The remaining 4,337 connected components are much smaller, with the second-largest containing only 66 vertices. This justifies a focus on the giant component.

\subsection*{Properties of the Entire Graph}
The giant component exhibits a diameter of 24 and a global clustering coefficient of 0.085, indicative of moderate clustering within a vast, interconnected network. The density is $2.458 \times 10^{-5}$. When compared to a random graph of the same size and density, the clustering coefficient of the giant component is significantly higher, as the random graph has an average clustering coefficient of only 0.000432. This stark difference underscores the collaborative tendencies among artists.

Conversely, a regular lattice of comparable size and density yields a clustering coefficient of 0.5, much higher than the observed network. However, the lattice has an astronomically higher diameter of 37,097, emphasizing its lack of shortcuts between distant vertices. The Spotify network's diameter is much closer to that of the random graph (20), reflecting the small-world nature of this collaboration network: high clustering relative to random graphs and a relatively short diameter compared to regular lattices.

\section*{A Look at the Top 0.1\% by Degree}
\begin{figure}[H]
    \centering
    \includegraphics[width=1\linewidth]{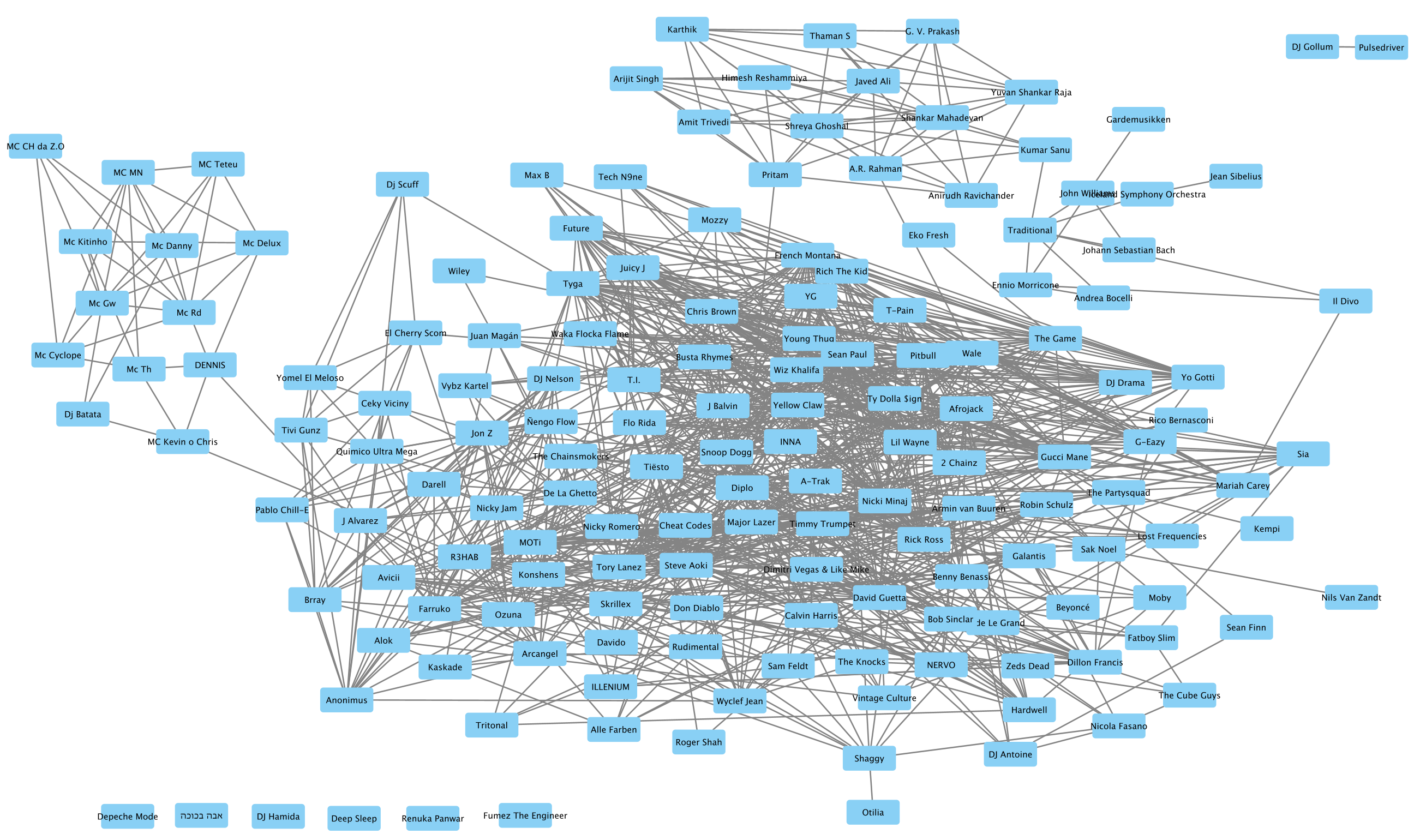}
    \caption{A graph of the top 0.1\% of artists based on number of connections.}
    \label{top_0.1}
\end{figure}

Visualizing the entire giant component in a single view is both computationally demanding and impractical due to its massive size. Figure~\ref{top_0.1} is a visualization of the top 0.1\% artists by vertex degree.

Top artists were much more likely to collaborate with other top artists, similar to what Park et al. observed in their network of classical composers \cite{P}. This phenomenon aligns directly with the power law distribution of the Spotify collaboration network and its scale-free nature, which will be discussed below. In the Spotify collaboration network, top artists play a central role in maintaining the network's connectivity and clustering.

\subsection*{Properties of the Chart-topper graph}
Since the dataset in question does not include any collaborative connections between the non-seed artists, this analysis will focus mostly on the 19,562 seed artists, which are those who made it to the Spotify weekly charts.

The chart-topper subgraph has a density of $3.77 \times 10^{-4}$. It exhibits a global clustering coefficient of 0.121, about ten times higher than the clustering coefficient of a comparable random graph (0.012). We can also observe that this clustering is higher than that of the entire graph of 156,326 artists, due to the mitigation of bias stemming from the omission of collaborations between non-seed artists. This absence leads to lower global clustering and a skewed degree distribution.

This elevated global clustering suggests that chart-topping artists form densely connected communities, likely driven by frequent and overlapping collaborations. The diameter of 18 for the chart-topper subgraph reflects the presence of efficient paths between artists. While this diameter is larger than that of a random graph (9), it is drastically smaller than the diameter of a comparable regular lattice (2,446). 

The small-world properties persist when examining the subgraph composed exclusively of these 19,562 chart-topping artists.
The remainder of this paper will focus almost entirely on this subgraph to eliminate the aforementioned bias.

\begin{figure}[H]
    \centering
    \includegraphics[width=1\linewidth]{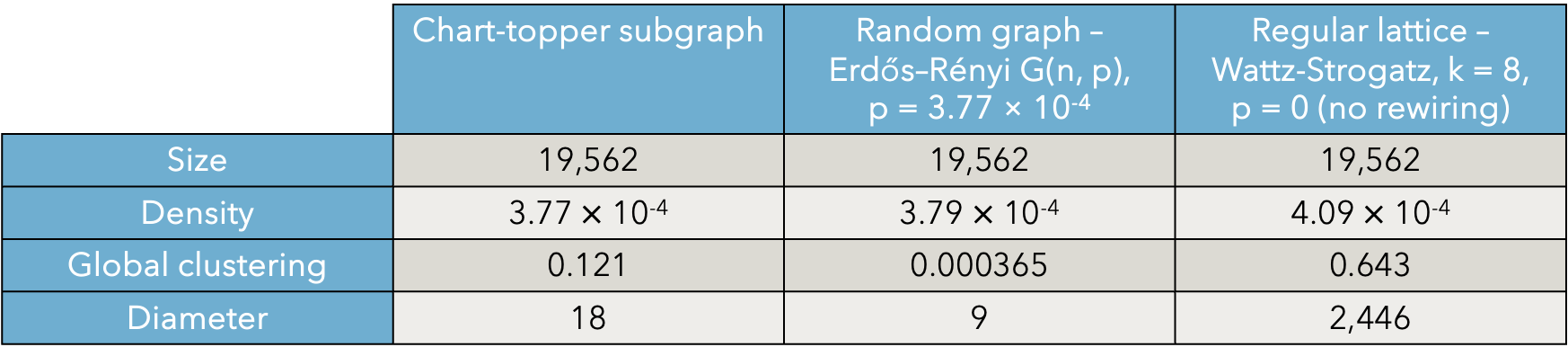}
    \caption{Comparison of network metrics for the chart-topper subgraph, a comparable random Erdős-Rényi graph, and a comparable regular lattice.}
    \label{chart_top}
\end{figure}

\subsection*{Vertex Degree Distribution}
\begin{figure}[H]
  \centering
  \begin{minipage}[b]{0.48\textwidth}
    \includegraphics[width=\textwidth]{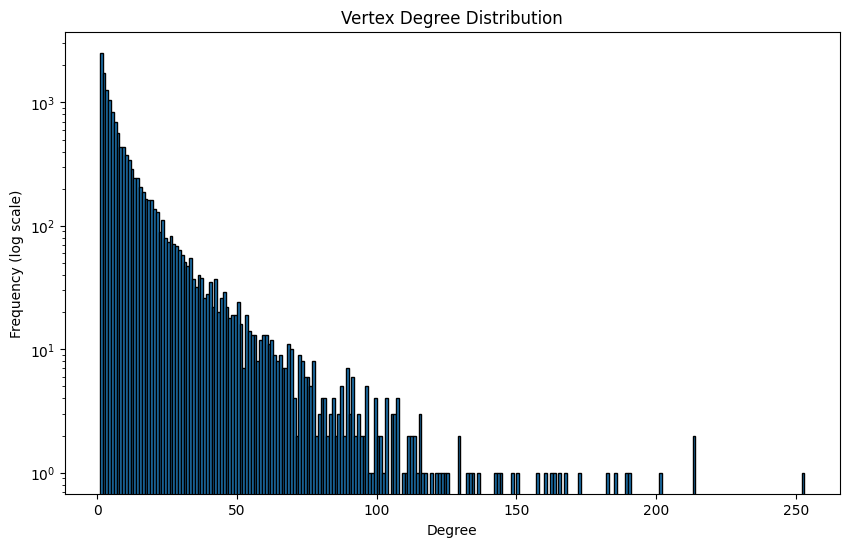}
    \caption{Histogram of the degree distribution.}
    \label{degrees}
  \end{minipage}
  \hspace{0.02\textwidth}
  \begin{minipage}[b]{0.48\textwidth}
    \includegraphics[width=\textwidth]{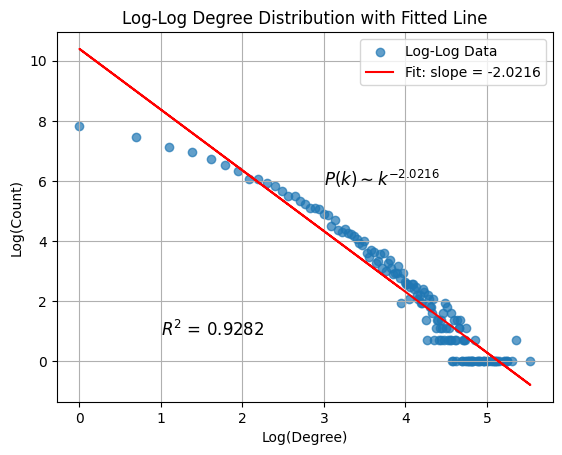}
    \caption{Log-log plot of the degree distribution with the fitted power law line.}
    \label{log-log}
  \end{minipage}
\end{figure}

The degree distribution of the chart-topper network follows a power law distribution, as characterized by 
\[
P(k) \sim k^{-2.0216}.
\]
The degree distribution follows a power law with $\gamma = 2.0216$. This fits within the lower end of the typical range of $2 < \gamma < 3$ for scale-free networks and suggests the dominance of highly connected hubs in the network, as the tail of the distribution is heavier. The fit of this power law is quite strong, with an $R^2$ value of $0.9282$.
Figures~\ref{degrees} and ~\ref{log-log} show the degree distribution and the power law fit.

\begin{figure}[H]
    \centering
    \includegraphics[width=1\linewidth]{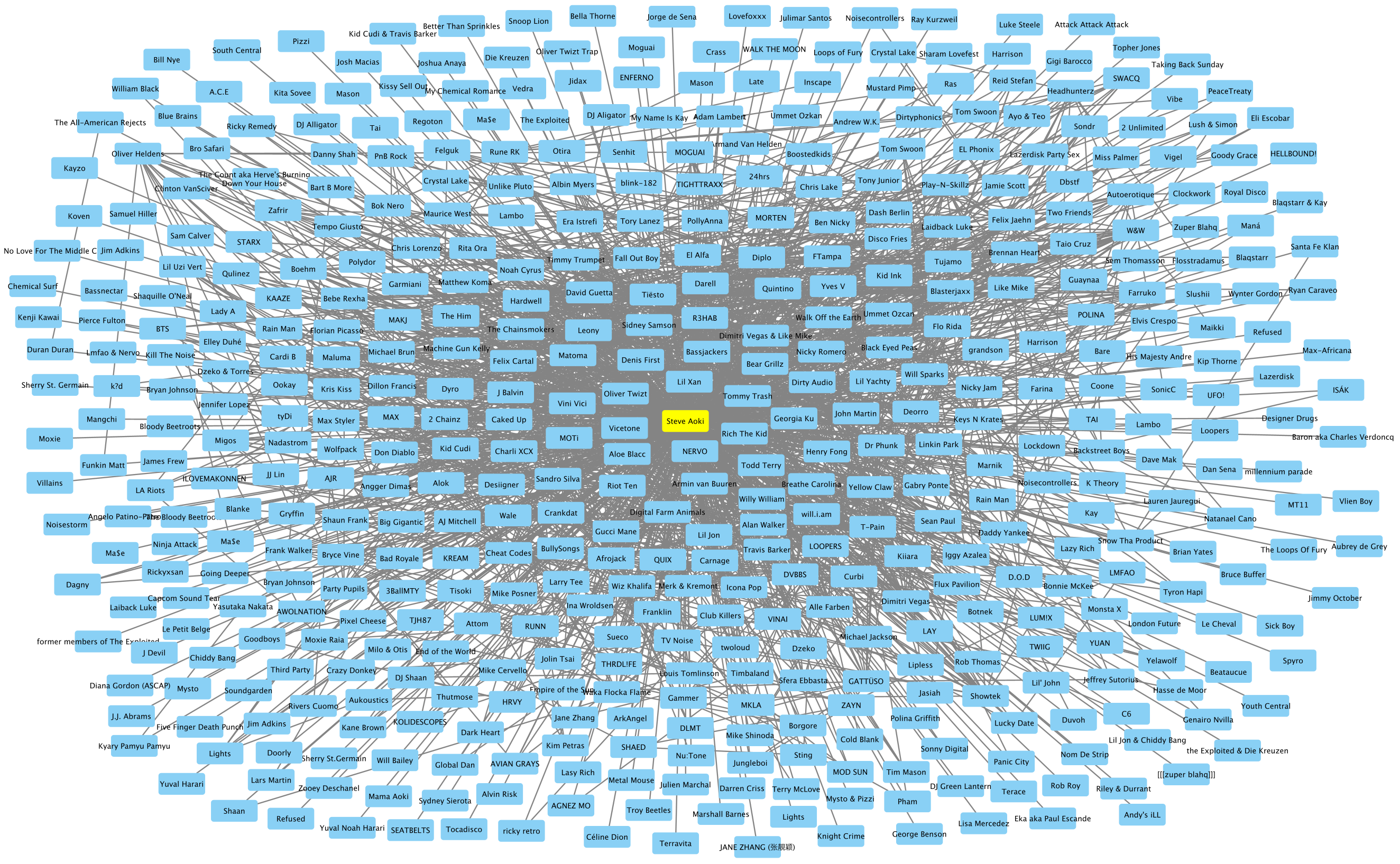}
    \caption{A visualization of Steve Aoki's (in yellow) connections and their connections with each other.}
    \label{steve}
\end{figure}

This distribution reflects the presence of a few highly connected vertices (hubs) in the network. 83 artists in the network have over 200 collaborations. The most connected artist, Steve Aoki, boasts 498 direct collaborative connections, including 23 connections among those artists with over 200 collaborations. Figure~\ref{steve} displays his entire graph of connections, which includes non-seed artists.

This pattern where popular artists gain connections faster over time aligns with Barabási and Albert's preferential attachment model \cite{Ba}, where nodes with higher degrees attract more connections over time, leading to the emergence of hubs. The presence of hubs like Steve Aoki further supports the scale-free nature, as they significantly reduce the average path length in the network, a hallmark feature of scale-free systems \cite{A}.

Hubs such as Aoki's play a pivotal role in maintaining the network’s small-world properties by providing pathways that reduce the average distance between vertices.

\section*{A Look at Chart-toppers by Country}
Following is an in-depth analysis of the artists who made the Spotify Weekly charts for the United States, the United Kingdom, Brazil, South Africa, India, and Japan. For each of these, subgraphs are created and small-world properties calculated and compared to those of random graphs and regular lattices of the same size and density.

\subsection*{United States Chart-toppers}
A narrower analysis focuses on a smaller subgraph of the 1,185 U.S. chart-topping artists and their 8,507 edges of collaboration. This subgraph presents even stronger small-world characteristics, with a clustering coefficient of 0.17188, which is higher than both the global clustering of the full chart-topper subgraph and the comparable random graph (0.0120). The diameter of this subgraph is remarkably small at 8, further emphasizing the efficiency of connections within this subset of artists. The comparable random graph has a diameter of 5, and the comparable regular lattice has a diameter of 85. These metrics highlight how the U.S. music industry fosters tightly-knit collaborations among its most successful artists, particularly those at the center of mainstream genres like rap, pop, and EDM.

\begin{figure}[H]
    \centering
    \includegraphics[width=1\linewidth]{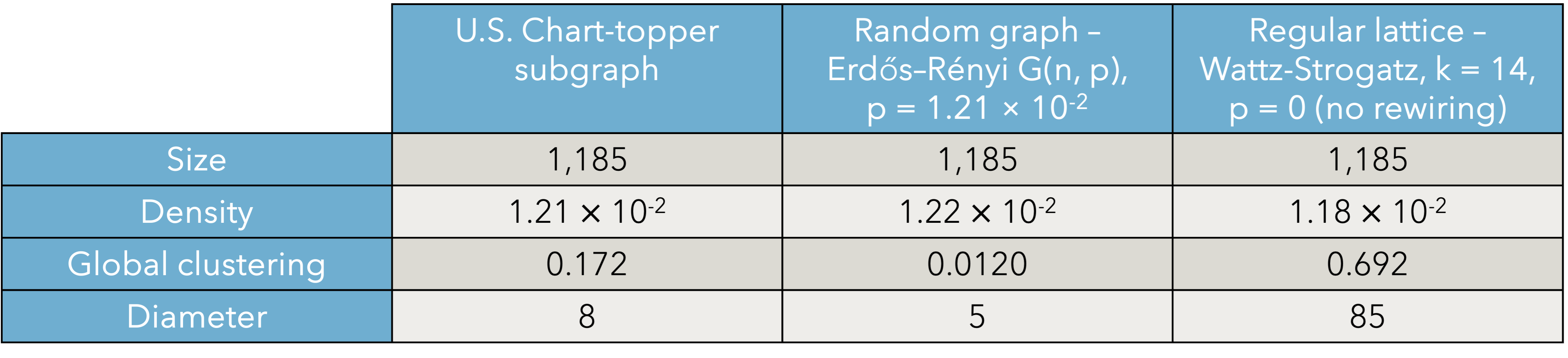}
    \caption{Comparison of network metrics for the United States chart-topper subgraph, a comparable random Erdős-Rényi graph, and a comparable regular lattice.}
    \label{chart_us}
\end{figure}

\subsection*{United Kingdom Chart-toppers}
The United Kingdom subgraph, consisting of 1,282 vertices and 8,131 edges, reflects a densely connected network with a global clustering coefficient of 0.145 and a diameter of 8. Compared to a random graph (clustering coefficient ~0.01 and diameter 5), the UK network displays substantially higher clustering, indicative of tight-knit collaboration patterns among chart-topping artists.  Its diameter is also much smaller than that of the comparable regular lattice (107). These metrics align with the UK's strong pop and alternative music scenes, where frequent collaborations between top artists like Ed Sheeran, Adele, and Clean Bandit reinforce genre-specific connectivity.

\begin{figure}[H]
    \centering
    \includegraphics[width=1\linewidth]{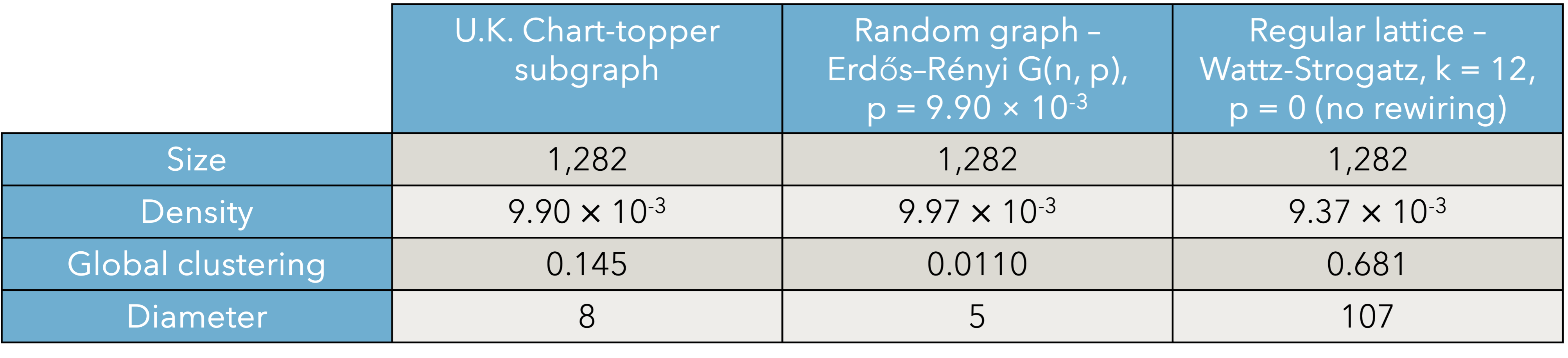}
    \caption{Comparison of network metrics for the United Kingdom chart-topper subgraph, a comparable random Erdős-Rényi graph, and a comparable regular lattice.}
    \label{chart_uk}
\end{figure}

\subsection*{Brazil Chart-toppers}
Brazil’s subgraph, featuring 1,081 vertices and 6,395 edges, demonstrates a clustering coefficient of 0.188 and a diameter of 8, highlighting a balance between local and global connectivity. The diameter is much lower than that of the comparable regular lattice (91), and the high clustering compared to that of same-size random graphs (0.0108) suggests cohesive communities within genres such as funk carioca, samba, and MPB (Música Popular Brasileira). These genres often foster localized hubs of collaboration, exemplified by artists like Anitta and MC Kevinho.

\begin{figure}[H]
    \centering
    \includegraphics[width=1\linewidth]{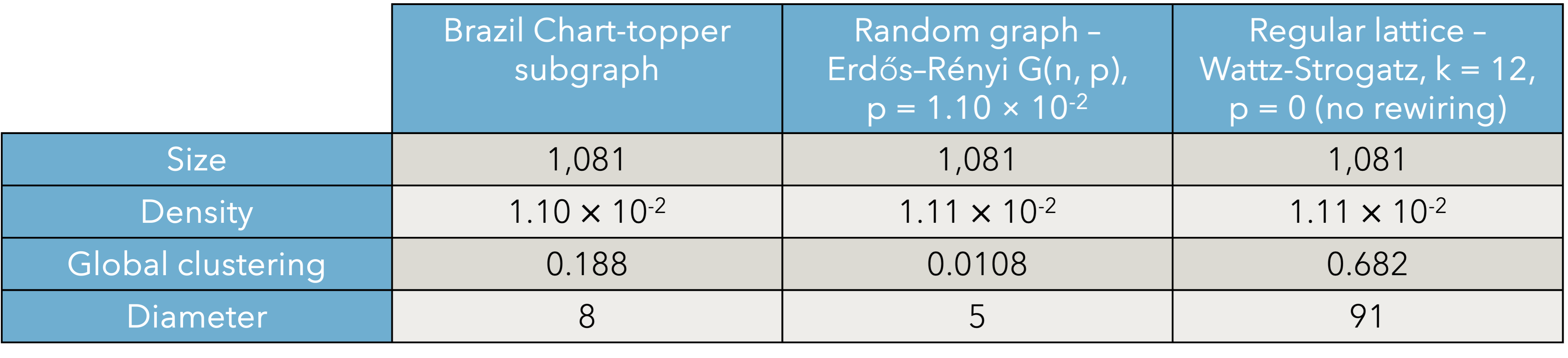}
    \caption{Comparison of network metrics for the Brazil chart-topper subgraph, a comparable random Erdős-Rényi graph, and a comparable regular lattice.}
    \label{chart_brazil}
\end{figure}

\subsection*{South Africa Chart-toppers}
The South African subgraph contains 746 vertices and 4,360 edges, with a global clustering coefficient of 0.206, the highest among the analyzed countries, reflecting the vibrant collaboration network in genres like Afrobeats, Amapiano, Kwaito, and Gospel. These genres are deeply rooted in South Africa’s cultural and musical heritage, often fostering tightly-knit artist communities that collaborate frequently within and across genres. Compared to that of the random graph (0.0154), the high clustering of the South African subgraph highlights the community-oriented nature of the industry. At 13, the diameter of the South Africa chart-toppers subgraph is relatively larger than that of the other countries, as it is further from that of the comparable random graph (5). The country's Gospel and Kwaito scenes promote recurring collaborations within religious and township music communities, further strengthening the network's cohesiveness. This strong clustering yet slightly elevated diameter supports the idea of localized yet globally impactful artist networks, with frequent partnerships driving connectivity.

\begin{figure}[H]
    \centering
    \includegraphics[width=1\linewidth]{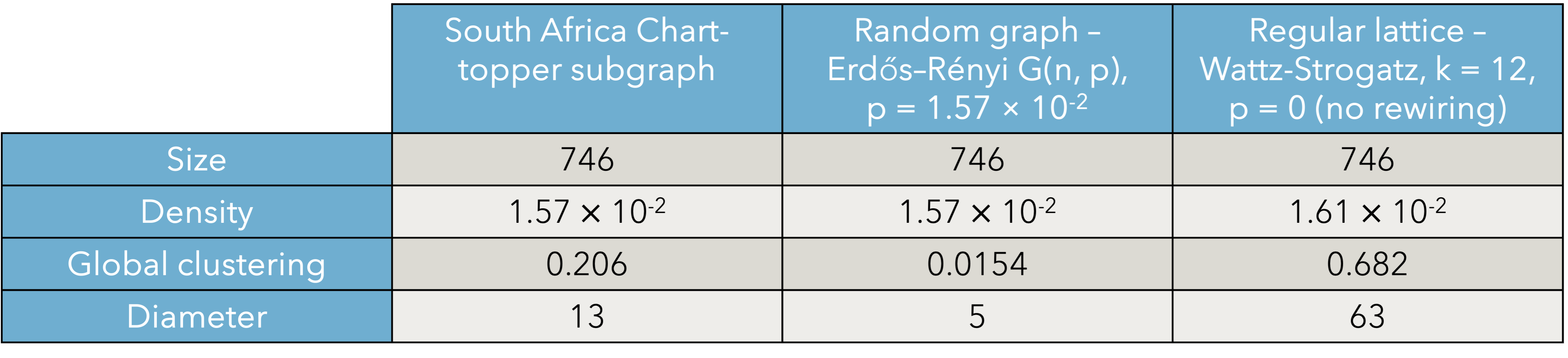}
    \caption{Comparison of network metrics for the South Africa chart-topper subgraph, a comparable random Erdős-Rényi graph, and a comparable regular lattice.}
    \label{chart_sa}
\end{figure}

\subsection*{India Chart-toppers}
India's subgraph, with 617 vertices and 3,019 edges, exhibits the second-highest clustering coefficient after South Africa at 0.203. The diameter of 9 is larger than most other regions, reflecting the network's geographic and linguistic diversity, but it is still not far off from that of the random graph (5), and it is much lower than that of the regular lattice (62). Bollywood and regional music industries dominate India's music scene, with collaborations often orchestrated by production houses, which act as central hubs. The clustering coefficient remains much higher than that of comparable random graphs (0.017), underscoring the cohesive community structure driven by these production-centered dynamics. The density of 0.0159 is also higher than other countries, indicating more frequent collaborations relative to the network size.

\begin{figure}[H]
    \centering
    \includegraphics[width=1\linewidth]{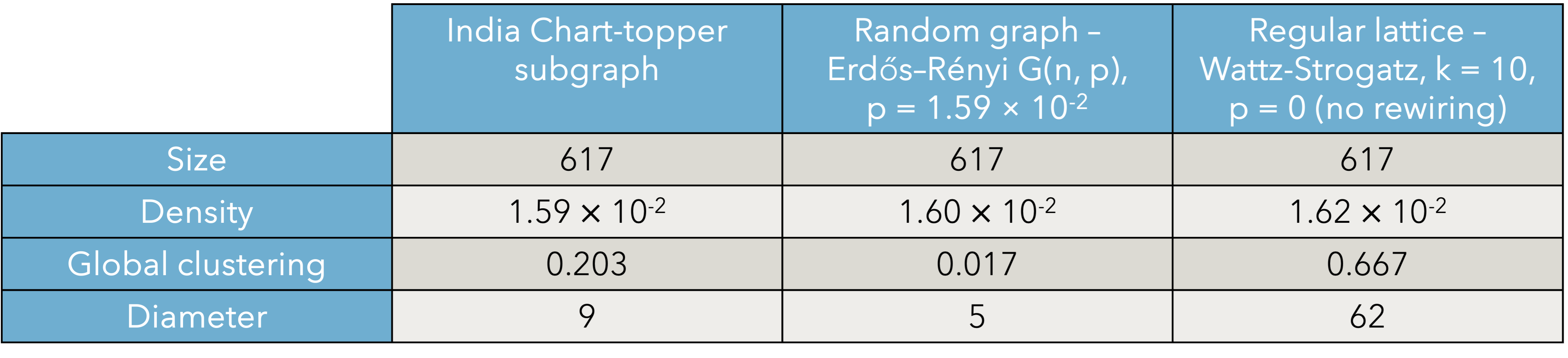}
    \caption{Comparison of network metrics for the India chart-topper subgraph, a comparable random Erdős-Rényi graph, and a comparable regular lattice.}
    \label{chart_india}
\end{figure}

\subsection*{Japan Chart-toppers}
The Japan subgraph consists of 677 vertices and 1,851 edges, with a global clustering coefficient of 0.104 and a diameter of 12. The Japanese music industry thrives on both local and global influences, and, like India, its diameter is slightly more elevated compared to that of the random graph (8). Still, the diameter of the comparable regular lattice is much higher at 113. The anime and gaming industries contribute significantly to the network, as artists like LiSA and Yoko Kanno collaborate on theme songs and scores that span various media. Genres like J-pop and anime scores dominate the network, fostering tightly-knit clusters of collaboration. Multiple artists collaborate frequently within highly organized production units. Compared to that of the comparable random graphs (0.00774), the clustering in Japan’s network highlights its unique production ecosystems, where artist collectives and multimedia franchises forge robust connections. This structure reflects the deeply interwoven nature of Japan’s music, entertainment, and cultural industries.

\begin{figure}[H]
    \centering
    \includegraphics[width=1\linewidth]{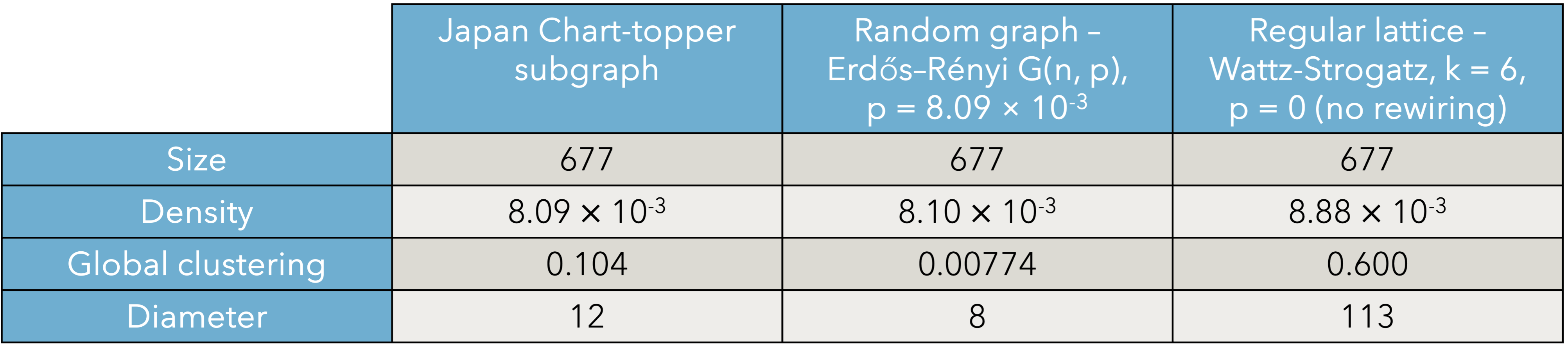}
    \caption{Comparison of network metrics for the Japan chart-topper subgraph, a comparable random Erdős-Rényi graph, and a comparable regular lattice.}
    \label{chart_japan}
\end{figure}

\subsection*{Significance of the Country-specific graphs}
The analysis of these subgraphs reveals significant insights into genre-specific and regional collaboration patterns and reinforces the importance of hubs and genre-based communities in shaping the network's structure.

All of these chart-topper subgraphs exhibit higher clustering coefficients and shorter diameters compared to random graphs, confirming the persistence of small-world properties across diverse cultural and regional music industries. The analysis highlights the universality of small-world traits in artist networks, driven by hubs and communities irrespective of cultural differences.

While diameters remain short, clustering varies more, with regions like India and South Africa showing stronger local cohesion, likely due to cultural or industry-specific factors. Regions with highly localized genres (e.g., India’s filmi music, Brazil’s funk carioca, and Japan's anime soundtracks) exhibit higher clustering, reflecting tight-knit collaborative communities. The United Kingdom, with more globally distributed music tastes, shows slightly lower clustering but retains cohesive structures due to frequent collaborations within its dominant genres. The United States falls in the middle, reflecting the dual nature of the United States music industry. Like the United Kingdom, the United States is a global hub for music, with collaborations often extending beyond national boundaries. South Africa, Japan, and India also exhibit slightly elevated diameters, likely due in part to the localized diversity of their music industries. This outward-looking aspect introduces some dispersion in collaborations, slightly lowering clustering compared to more localized networks. However, within the U.S., dominant genres like hip hop, pop, and country foster dense, localized collaboration networks. These genre-based communities contribute significantly to the clustering coefficient, as artists frequently feature one another, creating tight-knit subgroups.

\section*{Genre Distributions}
\subsection*{Genre Diversity Among All Artists}
\begin{figure}[H]
    \centering
    \includegraphics[width=1\linewidth]{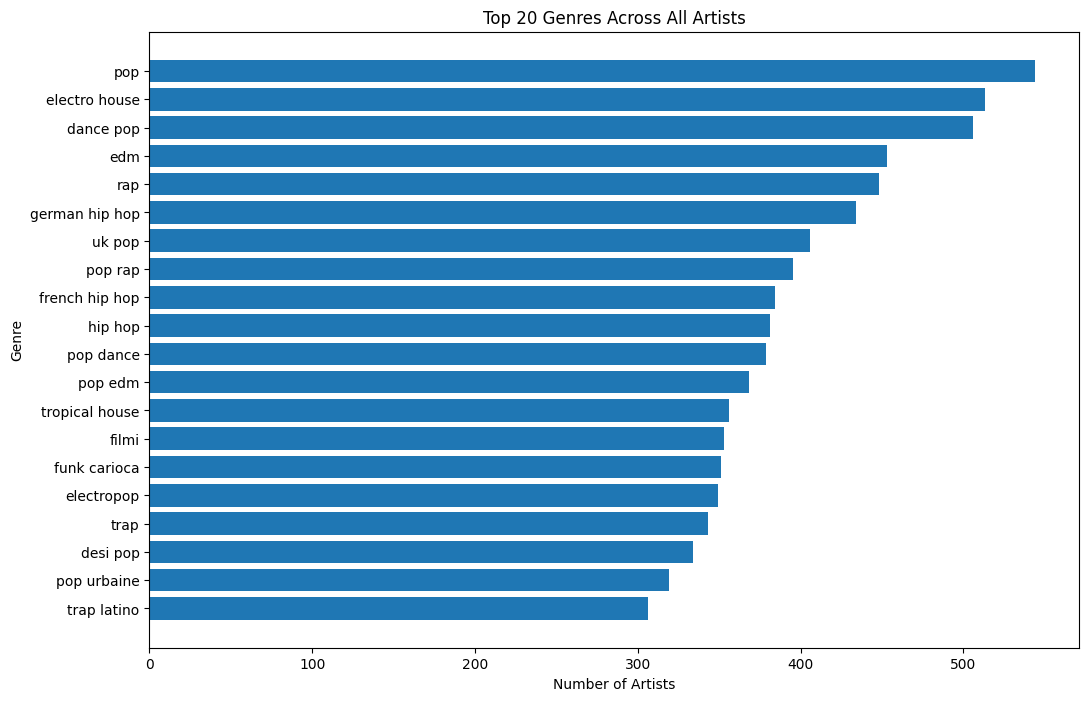}
    \caption{The top 20 genres among all artists in the Spotify dataset.}
    \label{all_genres}
\end{figure}

Figure~\ref{all_genres} highlights the broad spectrum of genres among all Spotify artists in the dataset. Notably, pop dominates with the highest number of artists, followed by electro house, dance pop, and rap. Genres such as tropical house, filmi, and funk carioca also appear, showcasing a global and diverse range of musical styles. This distribution reflects Spotify's inclusivity of global music cultures and the expansive nature of its artist base. Still, 102,133 artists have no genres associated with them in the dataset.

\subsection*{Genre Distribution Among Spotify Chart-Topping Artists}
\begin{figure}[H]
    \centering
    \includegraphics[width=1\linewidth]{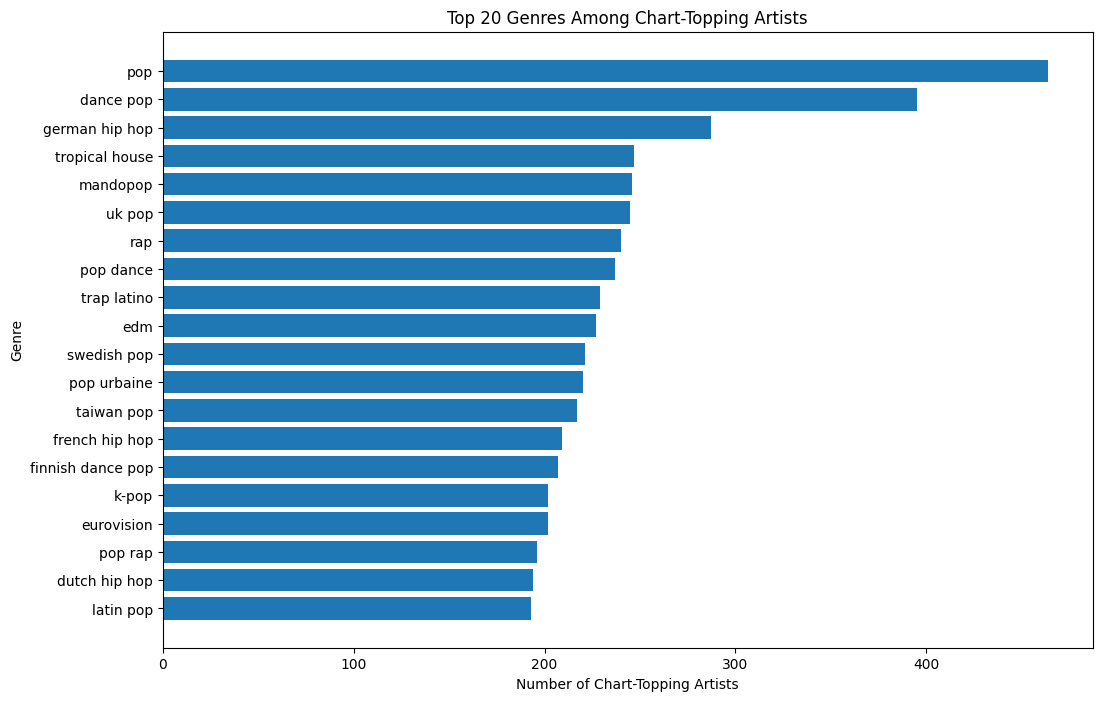}
    \caption{The top 20 genres among chart-topping artists in the Spotify dataset.}
    \label{chart_top_genres}
\end{figure}

In contrast, the Figure~\ref{chart_top_genres} indicates a shift toward more mainstream genres among the artists who have made it onto a Spotify weekly chart. Pop retains its dominance, but dance pop and German hip hop also rise to prominence, reflecting trends in popular global music consumption. Interestingly, genres like Mandopop and Swedish pop, which are less prevalent in the general artist population, gain significant representation among chart-toppers, suggesting that niche genres can achieve notable mainstream success within their target audiences.

\subsection*{Genre Trends Among U.S. Chart-Topping Artists}
\begin{figure}[H]
    \centering
    \includegraphics[width=1\linewidth]{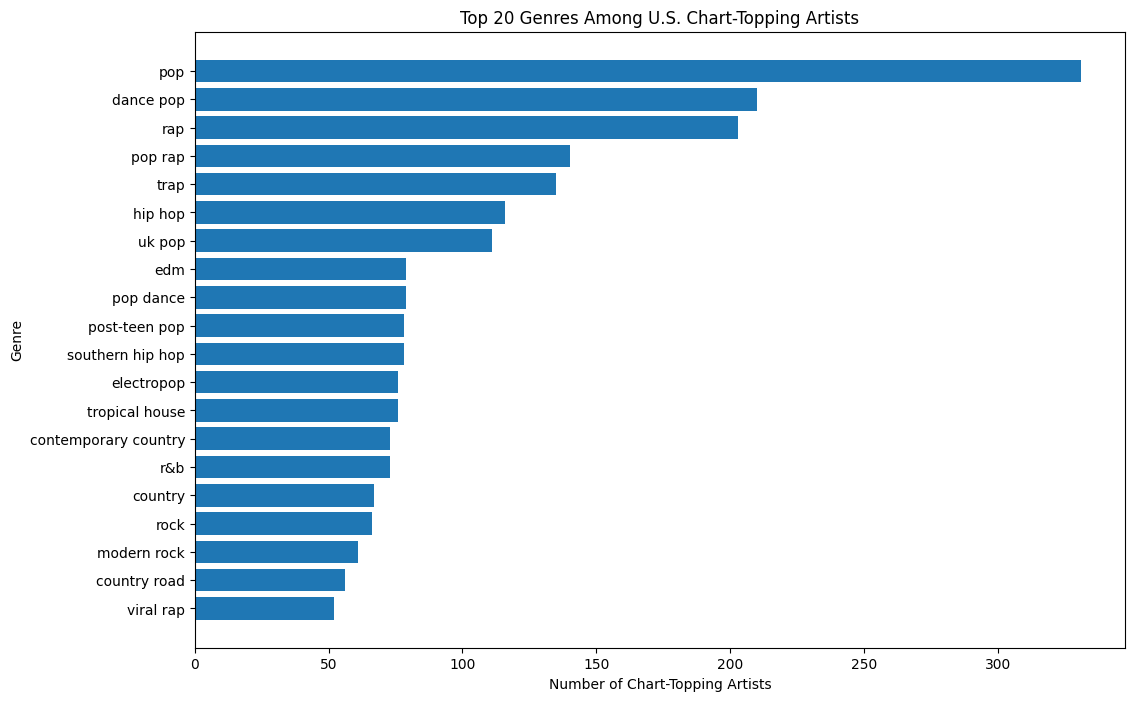}
    \caption{The top 20 genres among artists in the Spotify dataset who topped a U.S. weekly chart.}
    \label{us_genres}
\end{figure}

When focusing on the U.S. market, the genre distribution chart narrows further, emphasizing localized preferences. In Figure~\ref{us_genres}, it can be seen that pop and dance pop remain highly represented, but genres like rap, trap, and hip hop take on a more dominant role, reflecting the U.S.'s strong affinity for urban and hip-hop culture. Notably, genres like post-teen pop, country, and southern hip hop are unique to the U.S. chart-topping artists, showcasing regional trends distinct from the global chart-toppers.

It can be observed that pop remains a dominant genre across all groups, reflecting its universal appeal and broad listener base. Pop may be a wider genre within which many artists can easily fit.

However, distinct regional and global trends emerge when comparing chart-toppers. U.S. chart-toppers tend to favor urban and country genres, while global chart-toppers showcase a more diverse range of styles, including niche genres like Mandopop and Swedish pop. Interestingly, some genres, such as electro house, filmi, and funk carioca, which are prominent among all artists, fail to appear among chart-toppers, indicating their limited crossover success in global popularity charts. Despite this, regional genres like German hip hop and Mandopop underscore Spotify's ability to elevate non-English music to chart-topping status, emphasizing its role in promoting diversity and bridging cultural boundaries.

\section*{Genre Chart-toppers}
Genre-based chart-topper subgraphs reveal distinct collaboration patterns. Each subgraph is analyzed for its clustering coefficient, diameter, and density, compared with those of random graphs and regular lattices.

\subsection*{Dance Pop}
\begin{figure}[H]
    \centering
    \includegraphics[width=0.8\linewidth]{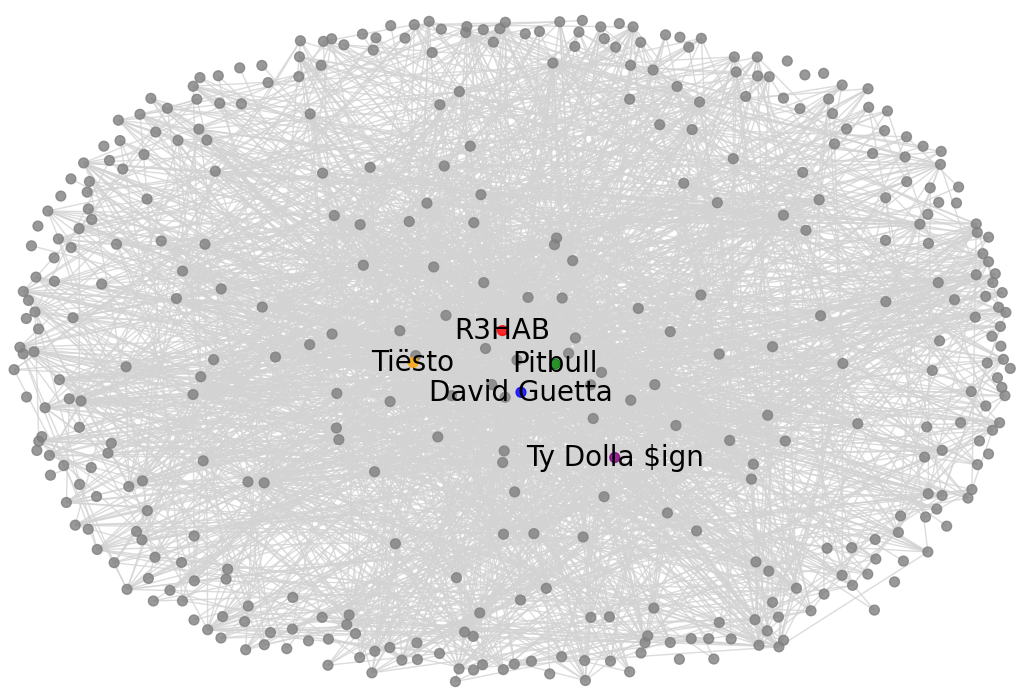}
    \caption{Dance Pop genre chart-topper subgraph with top 5 artists highlighted and labeled.}
    \label{dance_pop}
\end{figure}

There are 505 artists in this dataset associated with dance pop, and 355 of them are seed artists who have been on Spotify weekly charts. The subgraph of dance pop chart-toppers, shown in Figure~\ref{dance_pop}, has a global clustering coefficient of 0.192. When compared to random graphs of the same size and density, which have a clustering of approximately 0.0354, it can be seen that this genre demonstrates the highest clustering among the analyzed genres relative to the random graphs, reflecting a tightly-knit artist community. Additionally, the diameter of 6 is markedly lower than that of the comparable regular lattice (30). Hubs like R3HAB (90 connections, with 88 being chart-toppers) and David Guetta (87 connections, with 83 of them being chart-toppers) exemplify this interconnectedness.

\subsection*{Rap}
\begin{figure}[H]
    \centering
    \includegraphics[width=0.8\linewidth]{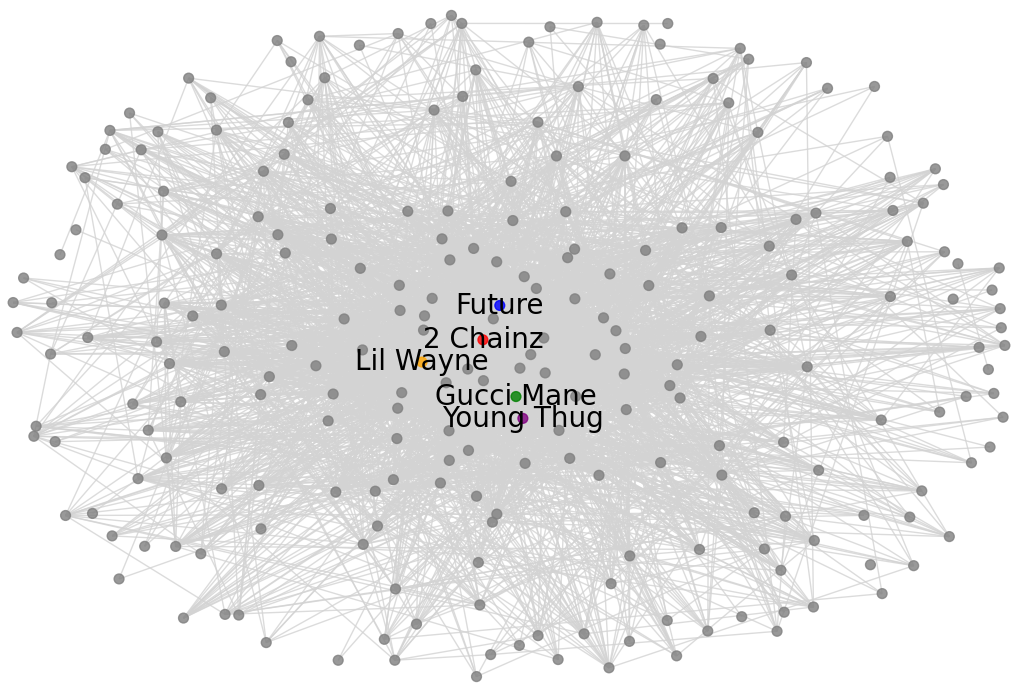}
    \caption{Rap genre chart-topper subgraph with top 5 artists highlighted and labeled.}
    \label{rap}
\end{figure}

The subgraph for rap chart-toppers, shown in Figure~\ref{rap} has the highest clustering coefficient (0.377) among analyzed genres by far, but the second-highest after dance pop relative to the comparable random graphs. Still, this illustrates strong interconnections between hub artists like Gucci Mane (130 connections, with 87 being chart-toppers) and Lil Wayne (101 connections, 84 being chart-toppers). In fact, this subgraph's global clustering is still over three times that of the comparable random graph (0.103). There are 448 artists in the dataset associated with this genre, but only 234 of them topped a Spotify chart. This demonstrates that many rap artists collaborate with chart-topping artists but are not chart-toppers themselves, perhaps reflecting the tendency for rap artists to be featured on the tracks of more mainstream pop artists. This is also seen in the relatively large gaps between the number of artists Gucci Mane and Lil Wayne have collaborated with versus how many of them are chart-toppers. Rappers, whether they are chart-toppers themselves or not, may be more likely to take on collaborations with less popular artists than other genres. The diameter of the rap chart-topper graph is 5, close to that of the comparable random graph (3), reflecting the genre’s collaborative culture and reliance on frequent partnerships. Meanwhile, the regular lattice with equal density and size has a significantly larger diameter of 10.

\subsection*{Tropical House}
\begin{figure}[H]
    \centering
    \includegraphics[width=0.8\linewidth]{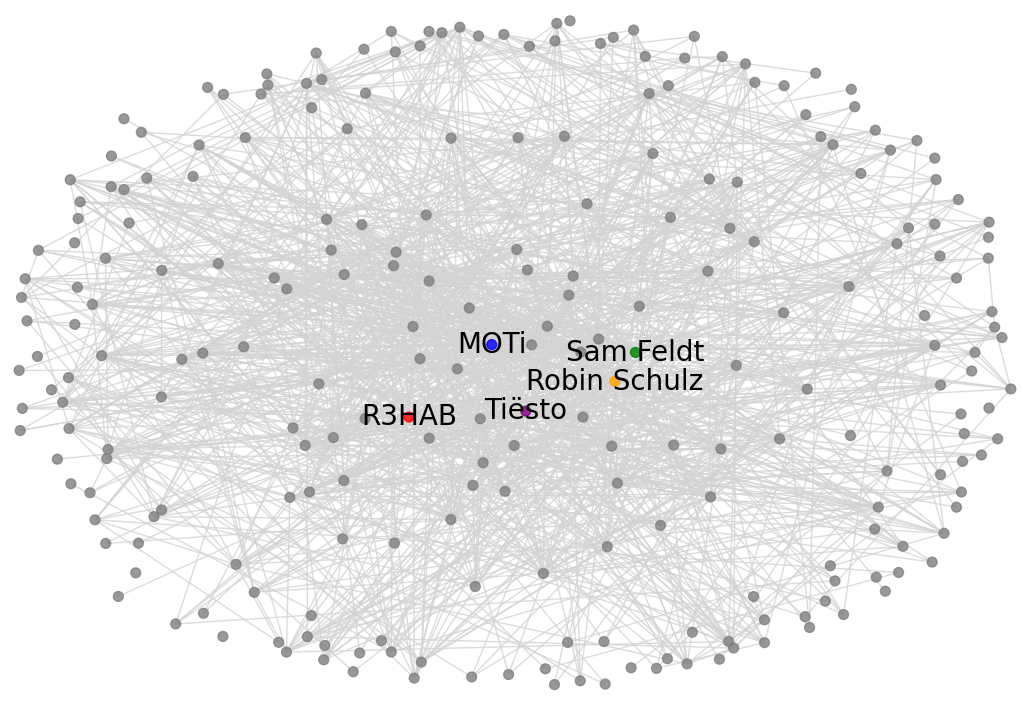}
    \caption{Tropical House genre chart-topper subgraph with top 5 artists highlighted and labeled.}
    \label{tropical_house}
\end{figure}

Tropical house has 355 associated artists in the complete graph, with 235 of them being chart-toppers. Its global clustering coefficient of 0.155, and it shows less interconnectivity than rap and even dance pop. However, it still shows strong connectivity when compared to the global clustering of a random graph of equal size and density (0.0427). This is driven by artists like Tiësto (56 connections, with 51 being chart-toppers) and R3HAB (47 connections, with 46 being chart-toppers). It is seen in Figure~\ref{tropical_house}.

\subsection*{Jazz}
The subgraph for jazz is surprising for a genre with such a highly collaborative history. Out of 74 total artists from the dataset associated with the genre, only 10 made up the giant component of the genre subgraph. Only 3 of these jazz artists made it onto a Spotify weekly chart. The two more highly connected jazz artists overall were Pat Metheney and Kurt Elling, with only 5 and 3 collaborations within jazz, respectively. This means that the vast majority of popular jazz artists on Spotify collaborate outside of the genre. This may spell trouble for the future of some regional genres such as jazz, as artists venture out of the genre to seek a wider, more globalized audience.

\subsection*{Implications for Small-World Validation}
These genre subgraphs consistently exhibit small-world properties. Their clustering coefficients are substantially higher than those of random graphs, while their diameters remain markedly smaller than those of regular lattices. The contrast is particularly pronounced in dance pop and rap, where high clustering aligns with close collaborative connections.

\section*{Genre Co-Occurrence}
A deeper examination of the entire Spotify collaboration network reveals significant genre co-occurrence patterns, where artists create music spanning multiple genres. These co-occurrences highlight not only the clustering within individual genres but also the interconnectivity across genres, forming a richly interconnected network.

Pairwise genre co-occurrences were counted across all artists, and a co-occurrence network was constructed, where vertices represent genres and edges represent significant co-occurrence counts (greater than 5). Using this network, the top genres by degree were identified. For each of those, the most frequently co-occurring genres were determined, along with their co-occurrence counts. This analysis is performed on the entire collection of artists since it only depends on artist data and not at all on collaborative connections.

Dance pop frequently co-occurs with genres like pop, post-teen pop, and tropical house. High co-occurrence counts, such as 236 with pop and 129 with post-teen pop, illustrate its centrality in the network.

Hip hop, an overarching culture of which subgenres rap, trap, and gangster rap are part, co-occurs often with pop rap, a hybrid genre. This serves as a bridge connecting rap with pop influences, co-occurring frequently with southern hip hop and gangster rap.

EDM and electro house exhibit significant overlap, with a high co-occurrence count of 302. This overlap reflects the fluid boundaries between these genres and is likely driven by shared production techniques and similar audience preferences.

\section*{Community Detection and Subgraph Properties}
\subsection*{Louvain Method and Community Dynamics}
The Louvain method for community detection is a heuristic algorithm that optimizes modularity, a measure of the quality of a network's partition into communities. It has two iterative phases. First, nodes are reassigned to neighboring communities if the move increases modularity. Second, the identified communities are aggregated into super-nodes to form a smaller network. These two steps are repeated until no further improvement in modularity is possible, resulting in a hierarchical structure of communities that reveals the network's organization at multiple levels. The method is computationally efficient, making it suitable for large-scale networks. The Louvain method outperforms other algorithms in both computational speed and the quality of the detected communities, as measured by modularity \cite{B}. 

Applying the Louvain method to the network of artists who made the Spotify weekly charts resulted in the detection of 5907 communities with a modularity score of 0.7780. A high modularity value (greater than 0.7) indicates well-separated communities, suggesting that artists within each detected community frequently collaborate with one another but have relatively fewer collaborations outside their community. This supports the hypothesis that artist collaborations are highly structured, likely reflecting genre-based, regional, or industry-driven collaborations. Due to the heuristic nature of the Louvain method, results may vary slightly across runs, but modularity consistently remains high (\textgreater 0.75), confirming the robustness of the detected communities. The analysis of these communities reveals several key insights into community dynamics.

\begin{figure}[H]
    \centering
    \includegraphics[width=0.7\linewidth]{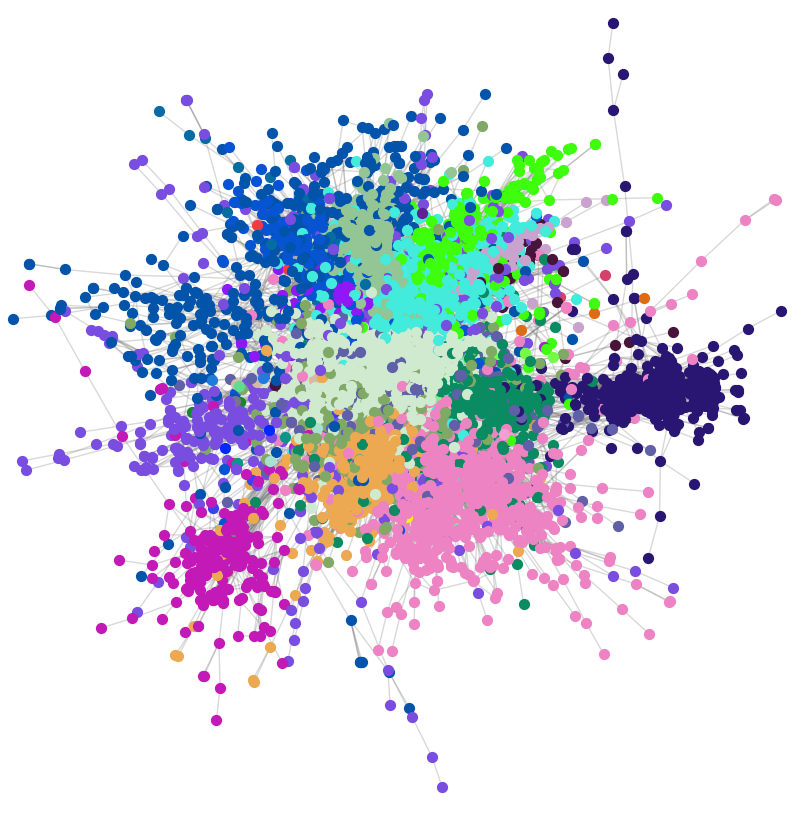}
    \caption{The top 10 communities identified by Louvain method for community detection.}
    \label{louvain_10}
\end{figure}

This method identifies distinct communities within the chart-topper subgraph, with clustering coefficients and diameters varying widely across communities. The top 10 of these communities by number of artists is seen in Figure~\ref{louvain_10}. The largest three of these communities are discussed below.

The largest community detected includes 1,928 vertices and represents a cluster of dance and pop artists who frequently collaborate. Notable artists include Diplo, Katy Perry, Billie Eilish, and David Guetta. This community has a diameter of 10, not far in comparison to the diameter of the random graph of equal size and density (6), however much lower than that of the comparable regular lattice (193). The global clustering is 0.111, which is nearly eighteen times higher than the random graph's (0.00269).

\begin{figure}[H]
  \centering
  \begin{minipage}[b]{0.48\textwidth}
    \includegraphics[width=\textwidth]{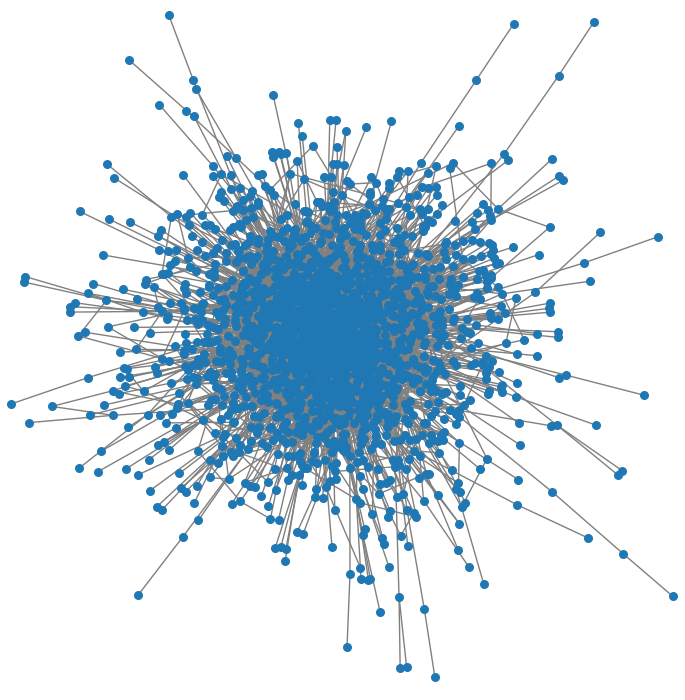}
    \caption{Graph of dance- and pop-dominated community.}
    \label{pop_louvain_G}
  \end{minipage}
  \hspace{0.02\textwidth}
  \begin{minipage}[b]{0.48\textwidth}
    \includegraphics[width=\textwidth]{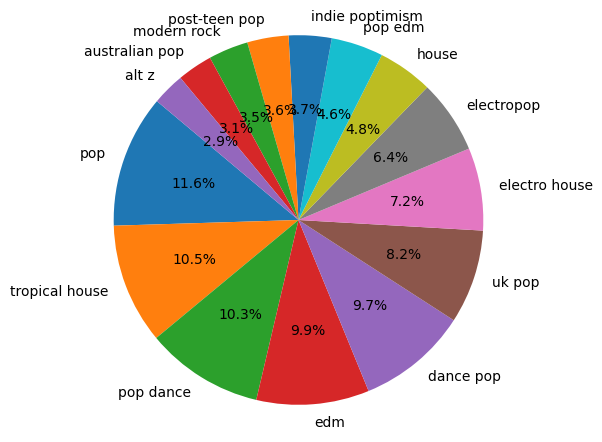}
    \caption{The genres found in the dance- and pop-dominated community.}
    \label{pop_louvain}
  \end{minipage}
\end{figure}

The second-largest community detected by the Louvain method is a group of 1,604 artists associated with modern Latin music, including trap Latino, reggaeton, and Latin pop. The graph representing this community has a diameter of 10. Again, this is higher than that of comparable random graphs (5) but much lower than that of the regular lattice (115). Additionally, the global clustering (0.233) is over 27 times that of the random graphs (0.00860). This community is driven by globally popular Latin American artists such as Bad Bunny, Ricky Martin, and Daddy Yankee.

\begin{figure}[H]
  \centering
  \begin{minipage}[b]{0.48\textwidth}
    \includegraphics[width=\textwidth]{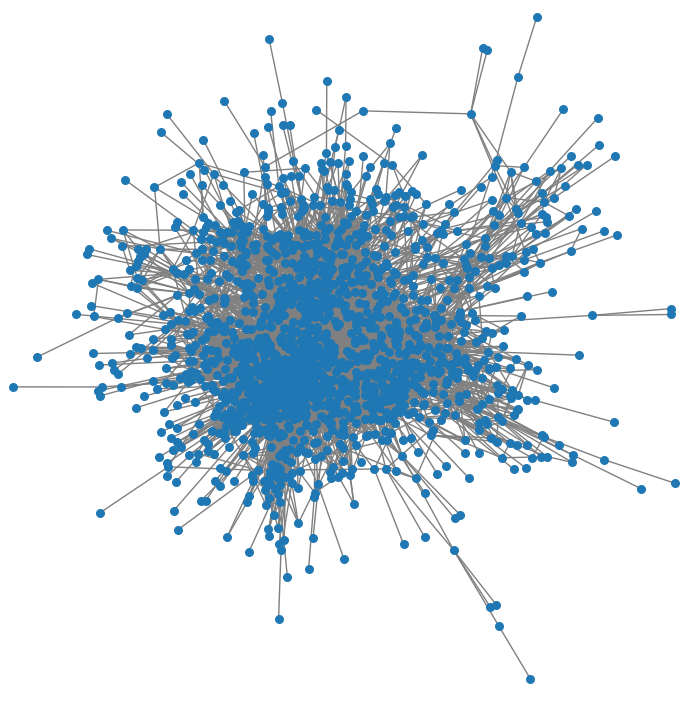}
    \caption{Graph of the Latin artist community.}
    \label{latinx_louvain_G}
  \end{minipage}
  \hspace{0.02\textwidth}
  \begin{minipage}[b]{0.48\textwidth}
    \includegraphics[width=\textwidth]{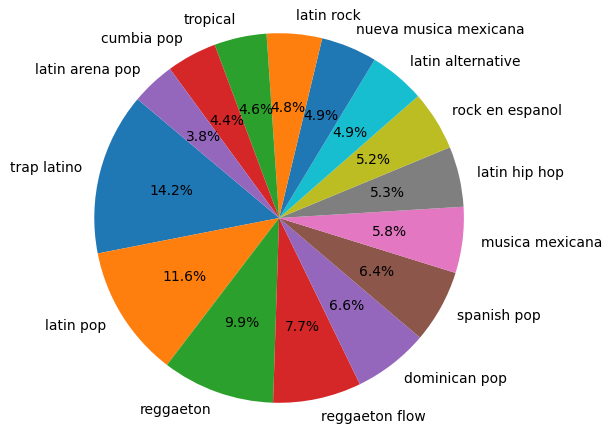}
    \caption{The genres found in this community characterized by Latin music.}
    \label{latinx_louvain}
  \end{minipage}
\end{figure}

The final Louvain-detected community to be discussed is one dominated by historically African American musical genres, including R\&B and especially hip hop and its subgenres. There are 1,165 artists in this group, and its graph has a diameter of 10. While higher than that of the comparable random graph (5), this is much lower than that of the regular lattice (84). At 0.197, the global clustering coefficient is over seventeen times higher than that of the random graph (0.0114). High clustering reflects the tight-knit nature of collaboration in these genres, driven by hubs such as Snoop Dogg, Gucci Mane, and Kendrick Lamar, and Beyoncé. Country is also represented in this mix of artists, which is unsurprising given the growing popularity of music that blends hip hop and country influences.

\begin{figure}[H]
  \centering
  \begin{minipage}[b]{0.48\textwidth}
    \includegraphics[width=\textwidth]{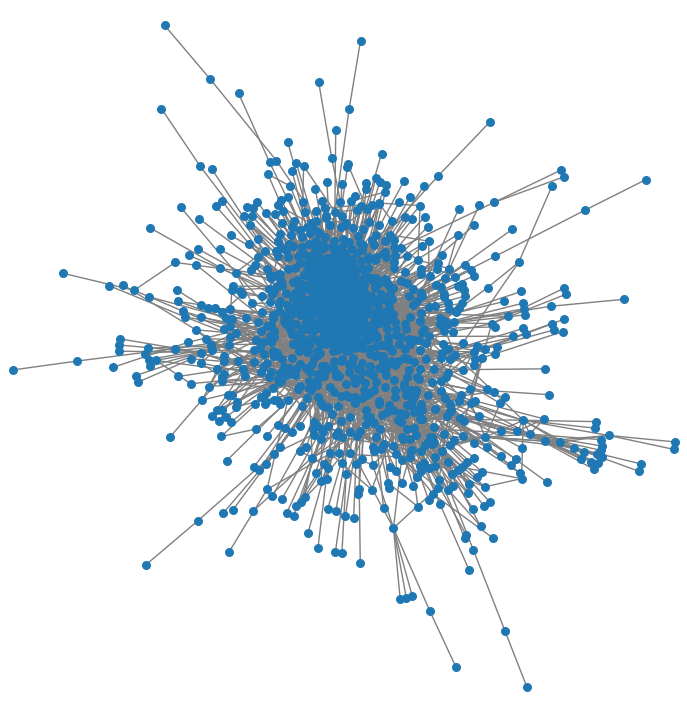}
    \caption{Graph of the hip hop-oriented community.}
    \label{aa_louvain_G}
  \end{minipage}
  \hspace{0.02\textwidth}
  \begin{minipage}[b]{0.48\textwidth}
    \includegraphics[width=\textwidth]{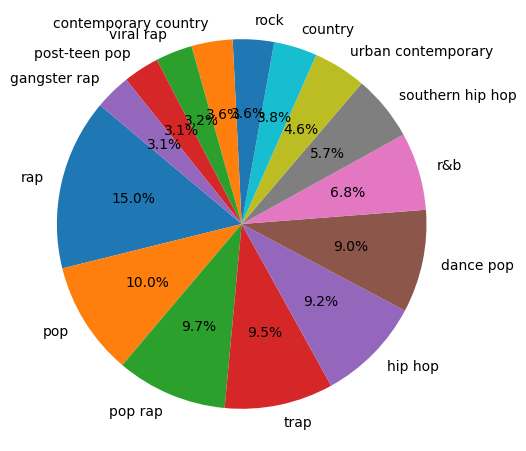}
    \caption{The genres found in this hip hop community.}
    \label{aa_louvain}
  \end{minipage}
\end{figure}

These community-level insights reinforce the network's modularity, demonstrating that artists tend to collaborate within genre or stylistic boundaries, with occasional cross-community interactions facilitated by high-degree vertex hubs.

\section*{Small-World Properties Validation}

\subsection*{Justifying Small-World Characteristics}
A network is considered a small-world network if it demonstrates:
\begin{enumerate}
    \item High clustering compared to a random graph.
    \item Low diameter compared to a regular lattice.
\end{enumerate}

Analyzing the collaboration network of Spotify artists who have topped a weekly chart, it is seen to exhibit:
\begin{itemize}
    \item{High clustering:} The global clustering coefficient is 0.121, far exceeding the clustering coefficient of a comparable random graph (0.000365). Genre subgraphs like rap exhibit even higher clustering, reaching 0.377.
    \item{Low diameter:} The network's diameter of 18 is much lower than the 2446 diameter of a comparable regular lattice, though slightly higher than the 9 for a random graph.
\end{itemize}

\begin{figure}[H]
    \centering
    \includegraphics[width=0.7\linewidth]{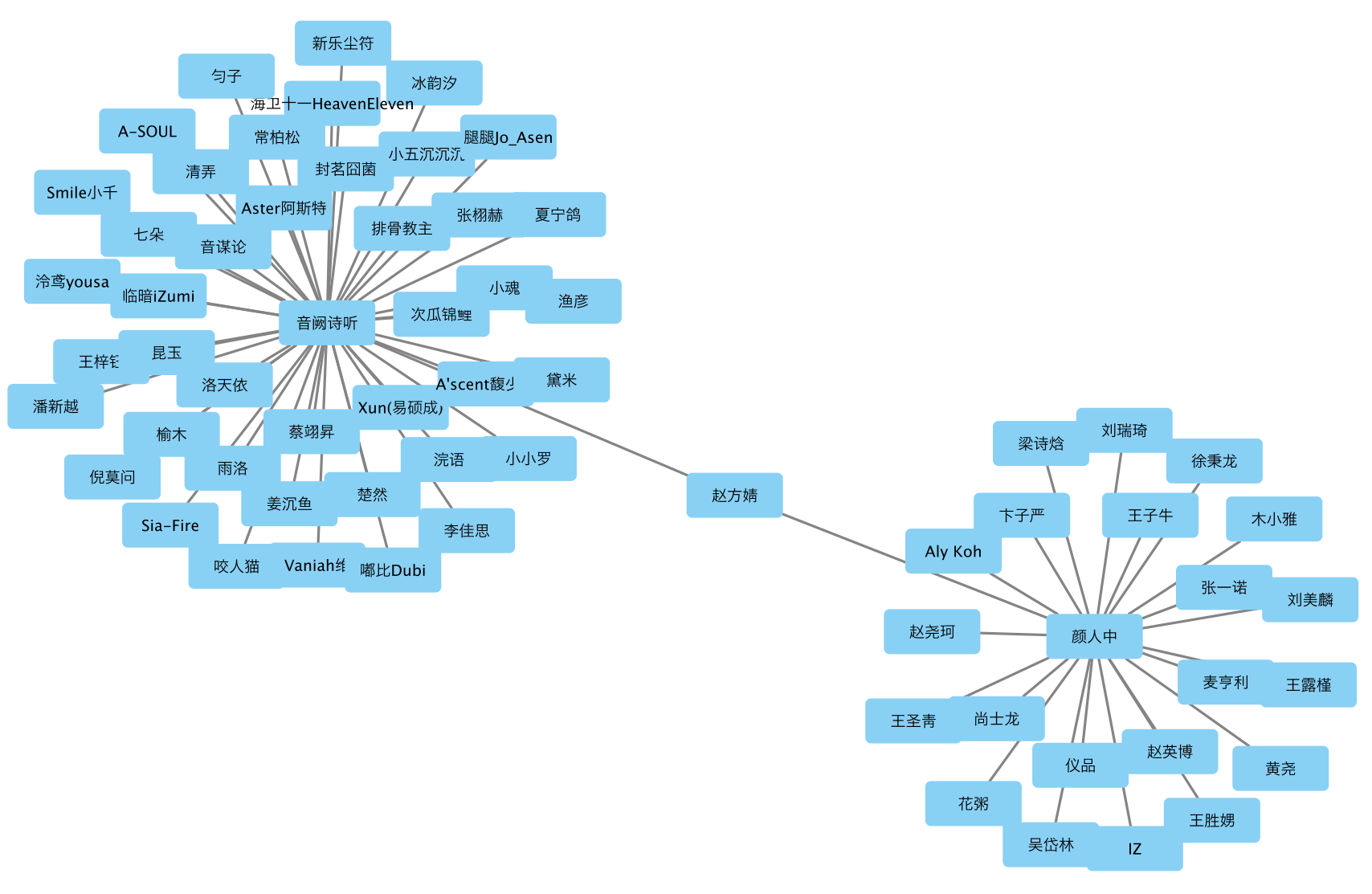}
    \caption{A bridge artist between two clusters of Chinese and Taiwanese artists.}
    \label{bridge}
\end{figure}

These properties confirm the small-world nature of the network. The presence of hubs, such as Steve Aoki's visualized earlier, reduces the average distance between vertices, while the clustering reflects strong local and genre-based collaborations. Additionally, bridge artists, such as the one in Figure~\ref{bridge}, play a crucial role in maintaining global connectivity across otherwise distinct communities. These artists act as key links through cross-genre and cross-community collaborations, which are essential for fostering connection among different clusters in the Spotify collaboration network, lowering diameter.

\subsection*{Implications of These Findings for Understanding Musical Collaboration}
The identification of small-world properties in the Spotify network has profound implications for understanding how artists collaborate. The small-world property holds across genres and regions. Some genres like rap demonstrate higher clustering, suggesting that artists within these communities work together more often, reinforcing genre identity. 

Highly connected artists bridge different subgraphs, facilitating cross-genre collaborations and promoting diversity in music creation. These genre-bending 'bridge' artists are vital to the global connectivity seen among artists on Spotify.

\section*{Limitations}
There are a few limitations in this study to consider.
Firstly, only the features of the original 19,562 chart-topping seed artists were scraped. Non-seed artists, despite having potential collaborations, are not connected to each other in this dataset, only with seed artists. This introduces a bias toward seed artists, potentially underestimating the clustering and connectivity of non-seed artists' subgraphs. However, this bias is mitigated by an analysis focused on the subgraph of chart-toppers, among whom all connections are recognized in the dataset. As expected, this graph displayed small-world properties to a stronger extent than the graph as a whole.

Another consideration is that the dataset treats collaborations as undirected edges, which simplifies the network by overlooking the potential dynamics of directed collaborations (e.g., primary vs. secondary artists on tracks).

The dataset's pre-defined genre categories may not fully capture cross-genre collaborations or subgenres, limiting the granularity of the analysis.

Additionally, while the dataset covers a period from 2013 to 2022, it lacks time stamps for collaborations, preventing an analysis of how collaboration trends evolve over time or how the network structure changes dynamically.

\section*{Applications}
There are several potential applications of these findings, including for music recommendation systems and music industry research.

Algorithms can leverage community structures and clustering coefficients to suggest collaborations or playlists tailored to user preferences. Understanding hub roles allows platforms like Spotify to recommend lesser-known artists connected to popular hubs using community-aware recommendation systems that leverage the Louvain method for community detection and small-world properties. Spotify currently uses machine learning techniques such as collaborative filtering, hybrid recommendation, and feedback loops to make its much-loved recommendations. User-based filtering, a part of collaborative filtering, relies on user similarities to make recommendations, similar to a network of users as nodes and similarities as weighted edges \cite{W}. However, perhaps the use of an artist collaboration network in conjunction with this could help improve user recommendations.

Clustering data can help identify emerging communities or rising stars within genres, providing actionable insights for record labels. Collaboration trends highlight which artists are central to specific communities, guiding promotional efforts. Predictive models could use degree distributions and clustering data to identify potential collaborations, increasing the likelihood of successful partnerships. Network properties could be correlated with financial success metrics, such as streaming revenue or concert sales.

\section*{Future Work}
In the future, it would be beneficial to extend the data scraping to include collaborations between non-seed artists to build a more complete network. Additionally, analyzing the evolution of the network over time to study how collaboration trends and community structures change would be beneficial.

Treating collaborations as directed edges to investigate whether certain artists consistently play supporting roles as song features could be rewarding. This may show, for example, that singers more often feature rappers than singers on their tracks.

\section*{Conclusion}
The analysis of this Spotify artist collaboration network provides compelling evidence of its small-world nature. By contrasting clustering coefficients and diameters of the entire graph and various subgraphs with those of comparable random graphs and regular lattices, this study confirms the network's small-world properties.

The network's power law degree distribution highlights the role of hubs in maintaining connectivity. Genre-based subgraphs and community detection emphasize the modularity of collaboration patterns. The small-world properties are shown to hold amongst both global and regional chart-topper subgraphs. The application of the Louvain algorithm revealed densely connected communities, underscoring the importance of genre and region in shaping artist networks.

These insights have far-reaching implications for the music industry, from improving recommendation systems to fostering innovation through targeted collaborations. Future research could explore how these network dynamics evolve over time or in response to industry changes, such as the rise of virtual collaborations.

\section*{Appendix A: Reproducibility and Code Availability}
The dataset used in this study is publicly available on \href{https://www.kaggle.com/datasets/jfreyberg/spotify-artist-feature-collaboration-network}{Kaggle}, and a cleaned version is hosted on my \href{https://github.com/raquelanamb/spotify-feature-network}{GitHub}.
All code and analyses are available in the GitHub repository.

For full replication steps, including dependencies and execution instructions, please refer to the README in the repository.

\section*{Appendix B: Supplemental Figures}
\begin{figure}[H]
    \centering
    \includegraphics[width=0.7\linewidth]{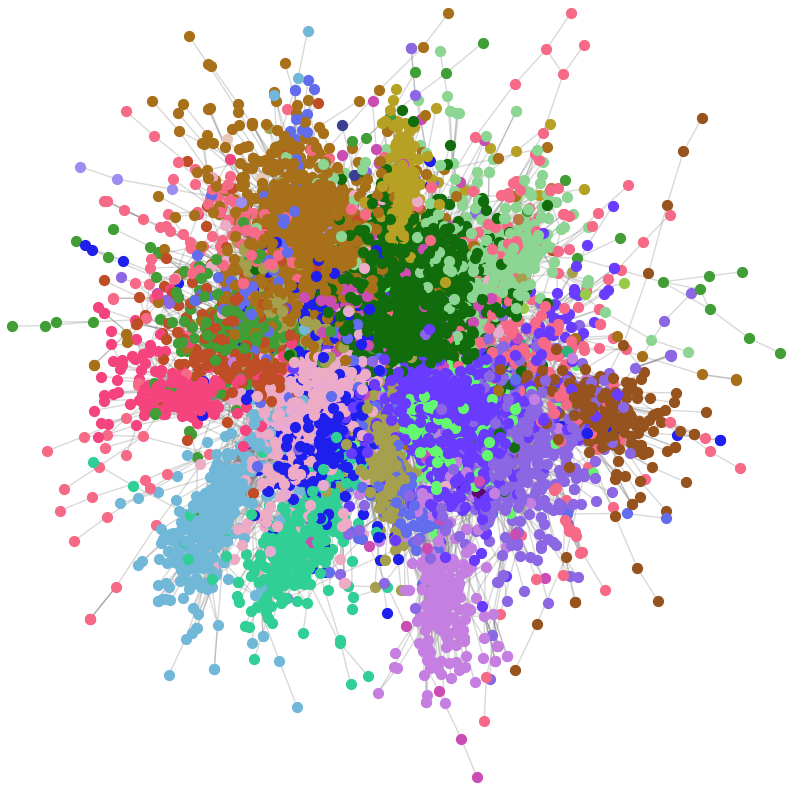}
    \caption{The top 20 communities identified by the Louvain method for community detection.}
    \label{louvain_20}
\end{figure}

\begin{figure}[H]
    \centering
    \includegraphics[width=0.9\linewidth]{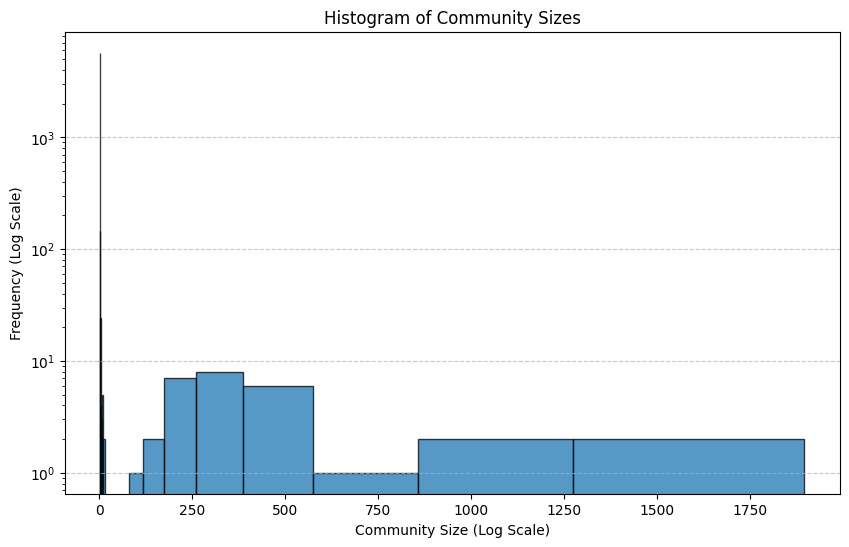}
    \caption{Distribution of sizes of all communities identified by the Louvain method for community detection. We can see that there are many minuscule communities and few large communities.}
    \label{comm_size_hist}
\end{figure}

The following are the three next-largest artist communities detected by the Louvain method for community detection.

\begin{figure}[H]
  \centering
  \begin{minipage}[b]{0.48\textwidth}
    \includegraphics[width=\textwidth]{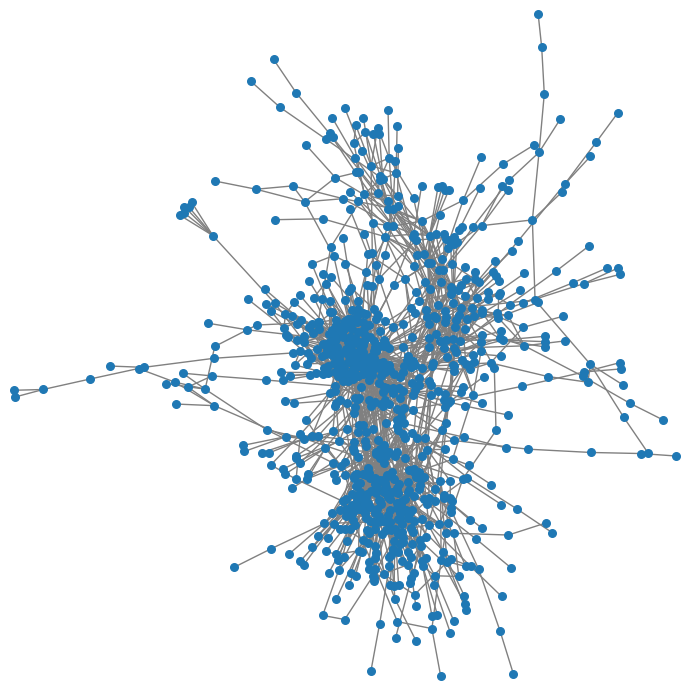}
    \caption{Graph of a predominantly Swedish and Norwegian pop and hip hop artist community.}
    \label{swedish_G}
  \end{minipage}
  \hspace{0.02\textwidth}
  \begin{minipage}[b]{0.48\textwidth}
    \includegraphics[width=\textwidth]{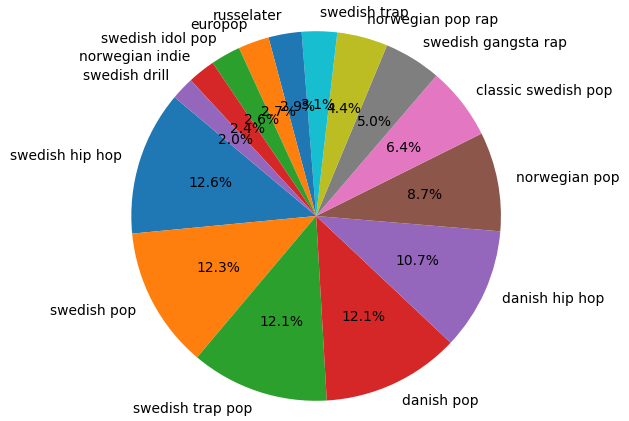}
    \caption{The genres found in this community characterized by Swedish and Norwegian pop and hip hop music.}
    \label{swedish}
  \end{minipage}
\end{figure}

\begin{figure}[H]
  \centering
  \begin{minipage}[b]{0.48\textwidth}
    \includegraphics[width=\textwidth]{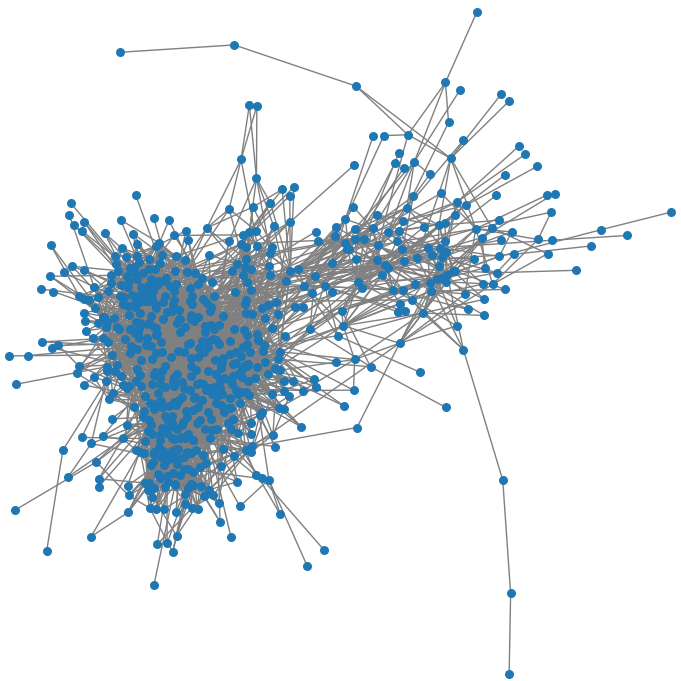}
    \caption{Graph of a predominantly Brazilian artist community, with high funk representation.}
    \label{brazilian_G}
  \end{minipage}
  \hspace{0.02\textwidth}
  \begin{minipage}[b]{0.48\textwidth}
    \includegraphics[width=\textwidth]{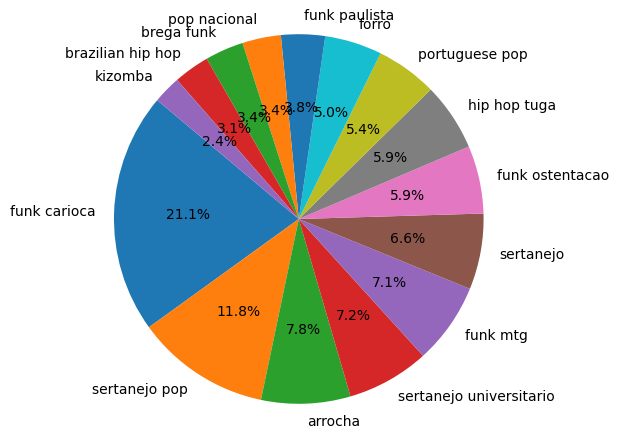}
    \caption{The genres found in this community characterized by Brazilian music.}
    \label{brazilian}
  \end{minipage}
\end{figure}

\begin{figure}[H]
  \centering
  \begin{minipage}[b]{0.48\textwidth}
    \includegraphics[width=\textwidth]{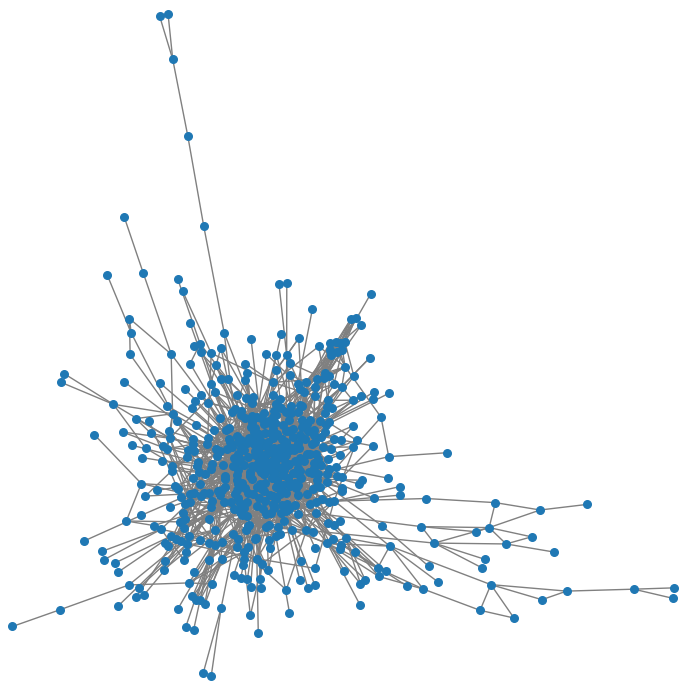}
    \caption{Graph of a predominantly German hip hop artist community.}
    \label{german_G}
  \end{minipage}
  \hspace{0.02\textwidth}
  \begin{minipage}[b]{0.48\textwidth}
    \includegraphics[width=\textwidth]{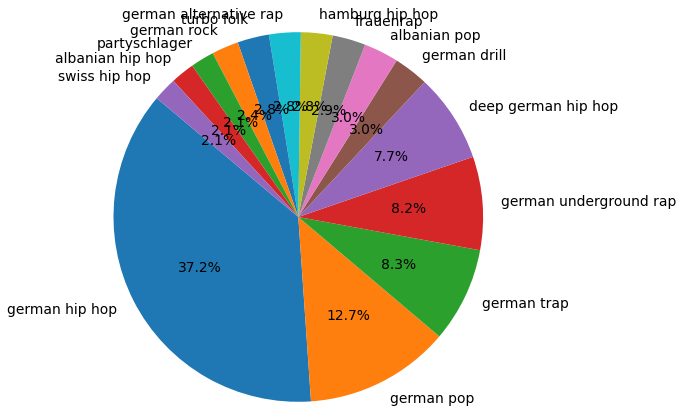}
    \caption{The genres found in this community characterized by German hip hop.}
    \label{german}
  \end{minipage}
\end{figure}

\section*{Acknowledgments}
The author thanks Professor Dana Fine of the University of Massachusetts Dartmouth for his guidance and valuable feedback during the development of this project.

\bibliographystyle{amsplain}

\end{document}